\documentstyle[prd,aps]{revtex}
\begin{document}
	
\title{Quantum Gravity of a Brane-like Universe
\thanks{Honorable mentioned, Gravity Research Foundation (1998)}}

\author{Aharon Davidson and David Karasik}
\address{Physics Department, Ben-Gurion University of the
Negev, Beer-Sheva 84105, Israel \\
(davidson@bgumail.bgu.ac.il, karasik@bgumail.bgu.ac.il)}

\maketitle

\begin{abstract}
Quantum gravity of a brane-like Universe is formulated, and its
Einstein limit is approached.
Regge-Teitelboim embedding of Arnowitt-Deser-Misner formalism,
parameterized by the coordinates $y^{A}(t,x^{i})$, is governed
by some $\rho_{AB}(y,y',y'')$.
Invoking a novel Lagrange multiplier $\lambda$, accompanying the 
lapse function $N$ and the shift vector $N^{i}$, we derive the
quadratic Hamiltonian
$$
{\cal H} = \frac{1}{2}N
\left[P_{A}\left((\rho-\lambda I)^{-1}\right)^{AB}P_{B} +
\lambda \right] + N^{i}y^{A}_{\,,i}P_{A} ~.
$$
The inclusion of matter resembles minimal coupling.
Setting $\displaystyle{P_{A}=-i\frac{\delta}{\delta y^{A}}}$, we
derive a bifurcated Wheeler-Dewitt-like equation.
Einstein gravity, associated with $\lambda$ being a certain
4-fold degenerate eigenvalue of $\rho_{AB}$, is characterized by
a vanishing center-of-mass momentum $\int_{}^{}P_{A}d^{3}x=0$.
Troublesome $(\rho-\lambda I)^{-1}$ is replaced then by regular
$M^{-1}$, such that $ M^{-1}(\rho-\lambda I)$ defines a projection
operator, modifying the Hamiltonian accordingly.
\end{abstract}

A prevailing theory is always seeded by a remarkably simple idea.
Regge-Teitelboim gravity\cite{RT}, a criticized rival\cite{DPR}
of Einstein gravity, may eventually fall into such a category.
After all, who can resist the philosophy that the first principle
which governs the evolution of the entire Universe is essentially
the one which determines the world-manifold behavior of particles,
strings and membranes.
Following such a viewpoint, the Universe, to be referred to as
a brane-like Universe, is viewed as a $4$-dim extended object\cite{GW}
floating in some (say) $10$-dim flat Minkowski background.
Some cosmological fingerprints\cite{D} of such a  brane-like
Universe have already been revealed.
Staying on practical grounds, however, Regge-Teitelboim gravity
needs not be considered a target by itself.
In fact, recalling its original underlying motivation, this theory
attempted to establish a viable mathematical trail towards the
unification of quantum mechanics with Einstein gravity.
This conjecture was driven by several remarkable facts:

\medskip
\noindent $\bullet$ Regge-Teitelboim gravity is, by construction, a
continuation of string theory. Unlike in Einstein gravity, the metric
tensor $g_{\mu\nu}(x)$ does not serve as a canonical field; this role
has been taken over by the embedding vector $y^{A}(x)$.
	
\noindent $\bullet$ Although Einstein equations are traded for
$[(G^{\mu\nu}-T^{\mu\nu})y^{M}_{\,;\mu}]_{;\nu}=0$, energy/momentum
conservation is still automatic.
	
\noindent $\bullet$ Regge-Teitelboim gravity exhibits a built-in
Einstein limit. In turn, every solution of Einstein equations is
automatically a solution of Regge-Teitelboim equations. 

\medskip
\noindent It has been speculated, relying on the structural similarity
to string/membrane theory, that quantum Regge-Teitelboim gravity may
be a somewhat easier task to achieve then quantum Einstein gravity.
The real target is then the Einstein limit of the theory, which in 
principle may call for additional first-class geometric constraints.
The trouble is, however, that the parent Regge-Teitelboim Hamiltonian
has never been derived!

\medskip
In this short essay, by deriving the quadratic Hamiltonian of a
gravitating brane-like Universe, we have overcome the dead-end reached
by Regge-Teitelboim, thereby opening the door for the quantum Einstein
gravity limit.
A key role in our formalism is played by a novel non-dynamical
field $\lambda$ which accompanies the standard Lagrange multipliers,
the lapse function $N$ and the shift vector $N^{i}$.
Starting from the purely gravitational case, the inclusion of
arbitrary matter serendipitously resembles minimal gauge coupling.
Altogether, the quantum theory prescribes a Virasoro-type momentum
constraint equation followed by a bifurcated Wheeler-Dewitt-like
equation.
Appealing to Poincare invariance of the embedding spacetime, a
generic Regge-Teitelboim configuration is parameterized by $\mu^{2}
>0$, recognized as the analogue of (mass)$^{2}$.
Quite surprisingly, an Einstein configuration turns out to be
characterized by $\mu^{2}= 0$.
In this language, Einstein gravity can be interpreted as the 'massless'
limit of Regge-Teitelboim gravity.

\medskip
Given the background Minkowski metric $\eta_{AB}$ and some
embedding vector $y^{A}(t,x^{i})$, the induced $4$-dim
line-element can be put in the Arnowitt-Deser-Misner (ADM)
form
\begin{equation}
	ds^{2}= -N^{2}dt^{2}+
	h_{ij}(dx^{i}+N^{i}dt)(dx^{j}+N^{j}dt) ~,
\end{equation}
provided the $3$-metric $h_{ij}$, the shift vector $N_{i}$, and
the lapse function $N$ are identified with
\begin{equation}
	h_{ij} = \eta_{AB}y^{A}_{\,,i}y^{B}_{\,,j} ~,~~
	N_{i} = \eta_{AB}y^{A}_{\,,i}\dot{y}^{B} ~,~~
	N^{2} = N_{i}N^{i}-\eta_{AB}\dot{y}^{A}\dot{y}^{B} ~.
\end{equation}
Notice the time-like unit vector ${\displaystyle n^{A}\equiv
\frac{1}{N}\left(\dot{y}^{A}-N^{i}y^{A}_{\,,i}\right)}$
orthogonal to $y^{A}_{\,,i}$.

\medskip
The gravitational Regge-Teitelboim Lagrangian density is the
standard one (the canonical fields are not).
Up to a surface term, it can be written in the form
\begin{equation}
	{\cal L} = -\sqrt{h}\left[ N{\cal R}^{(3)} -
	\frac{1}{N}({\cal K}_{ij}{\cal K}^{ij}-{\cal K}^{2}) +
	2N\Lambda \right] ~,
\end{equation}
where ${\cal R}^{(3)}$ denotes the 3-dim Ricci scalar constructed by
means of the 3-metric $h_{ij}$, and ${\cal K}_{ij}\equiv NK_{ij}$ is
the extrinsic curvature $K_{ij}$ factorized by the lapse function $N$.
${\cal K}_{ij}$ is free of mixed derivative $\dot{y}^{A}_{\, ,i}$-terms,
and since $\ddot{y}^{A}$-terms are absent in the first place, the Lagrangian
${\cal L}(y,\dot{y},y_{|i},y_{|ij},\ldots)$ is apparently ripe for the
Hamiltonian formalism.

\medskip
The fact that the $3$-metric $h_{ij}$ is $\dot{y}^{A}$-independent
helps us to derive the momenta $P_{A}$ conjugate to $y^{A}$, that is
\begin{equation}
	P_{A} \equiv \frac{\delta {\cal L}}{\delta \dot{y}^{A}} =
	\sqrt{h}\left\{\left[{\cal R}^{(3)}+
	\frac{1}{N^{2}}({\cal K}_{ij}{\cal K}^{ij}-{\cal K}^{2})
	+2\Lambda \right]n^{A}+ \frac{2}{N}({\cal K}^{ij}-
	h^{ij}{\cal K})y^{A}_{\, |ij}\right\} ~.
\end{equation}
To simplify the algebraic structure of $P^{A}$, define the
$\dot{y}^{A}$-independent tensor
\begin{equation}
	\rho^{AB} \equiv 2\sqrt{h}\left[
	(h^{ia}h^{jb}-h^{ij}h^{ab})y^{A}_{\,|ab}y^{B}_{\,|ij}+
	\left({\cal R}^{(3)}+2\Lambda\right)\eta^{AB}\right] ~,
\end{equation}
to finally arrive at
\begin{equation}
	\fbox{${\displaystyle P^{A}=\frac{1}{2}(n\rho n)n^{A}+
	\rho^{A}_{\,B}n^{B}}$}
\end{equation}

\medskip 
One can immediately verify, in analogy with Wheeler-DeWitt theory
and string theory, that the Hamiltonian ${\cal H}$ vanishes
\begin{equation}
	{\cal H} = \dot{y}^{A}P_{A}-{\cal L}=
	N\left(n^{A}P_{A}-\frac{1}{N}{\cal L}\right) +
	N^{i}y^{A}_{\,,i}P_{A} = 0 ~,
\end{equation}
and thus can be interpreted as a sum of constraints.
Invoking the powerful embedding identity $\eta_{AB}y^{A}_{|ij}y^{B}
_{\,,k} \equiv 0$, the first constraint $y^{A}_{\,,i}P_{A} = 0$
is easily extracted, reflecting the fact that $y^{A}_{\,,i}n_{A}=0$.
The second constraint is hidden within $\displaystyle{n^{A}
P_{A}-\frac{1}{N}{\cal L} = 0}$.
A naive attempt to solve $n^{A}(\rho, P)$ and substitute into
$n^{2}+1=0$, falls short.
The cubic equation involved rarely admits simple solutions, and
even in cases it does, the resulting constraint is anything but
a quadratic form in the momenta.

\medskip
The way out involves the definition of a quantity $\lambda$,
such that
\begin{equation}
	\fbox{$P^{A}=(\rho -\lambda I)^{A}_{\,B}n^{B}$}
\end{equation}
The price for an \textit{independent} $\lambda$ being an
additional constraint $n\rho n +2\lambda = 0$.
Assuming that $\lambda$ is {\it not} an eigenvalue of $\rho^{A}
_{\,B}$, we can solve for $n^{A}(\rho,P,\lambda)$ and find
\begin{equation}
	n^{A}=\left[\left(\rho-\lambda 
	I\right)^{-1}\right]^{A}_{\,B}P^{B} ~.
\end{equation}
The leftover constraints can then be grouped into
\begin{equation}
	P(\rho-\lambda I)^{-2}P +1 = 0 ~,~~
	P(\rho-\lambda I)^{-1}P +\lambda = 0 ~.
\end{equation}
The first of which, owing to ${\displaystyle \frac{d}{d\lambda}
(\rho-\lambda I)^{-1}= (\rho-\lambda I)^{-2}}$, can be regarded
superfluous provided we elevate $\lambda$ to the level of a
canonical non-dynamical variable.
Note in passing that the special case $\rho^{A}_{\,B}\sim\delta
^{A}_{\,B}$ corresponds to a Nambu-Goto string.
Explicitly, $\rho=4\Lambda\sqrt{h}I$ fixes $\lambda=2\Lambda\sqrt{h}$,
and gives rise to the familiar Virasoro constraint
$P^{2}+4\Lambda^{2}\eta_{AB}y^{A}_{,\sigma}y^{B}_{,\sigma}=0$.

\medskip
Altogether, the Regge-Teitelboim Hamiltonian acquires the
\textit{quadratic} form
\begin{equation}
  	\fbox{${\displaystyle {\cal H}=\frac{1}{2}N
  	\left[P_{A}\left((\rho-\lambda I)^{-1}\right)^{AB}P_{B}+
  	\lambda \right]+N^{i}y^{A}_{\,,i}P_{A}}$}
\end{equation}
with $N$, $N^{i}$, and notably $\lambda$ serving as Lagrange
multipliers.
$(\rho-\lambda I)^{-1}$ plays a role analogous to the
Wheeler-DeWitt metric on superspace.
Here, however, superspace has been traded for the embedding
spacetime itself, and $(\rho-\lambda I)^{-1}_{AB}$ needs not be
confused with the metric $\eta_{AB}$.
Once matter is included, the momenta $P_{A}$ conjugate to
$y^{A}$ receives an extra contribution
${\displaystyle \Delta P_{A}=\frac{\delta{\cal L}_{matter}}
{\delta \dot{y}^{A}}=\frac{1}{2}\sqrt{h}NT^{\mu\nu}
\frac{\delta g_{\mu\nu}}{\delta \dot{y}^{A}}}$.
Using the notations
\begin{equation}
	T_{nn} \equiv \left(T^{\mu\nu}y^{A}_{\,,\mu}y^{B}_{\,,\nu}\right)
	n_{A}n_{B} ~,~~
	T_{ni} \equiv \left(T^{\mu\nu}y^{A}_{\,,\mu}y^{B}_{\,,\nu}\right)
	n_{A}y_{B,i} ~,
\end{equation}
and bearing in mind that $T_{nn}(h_{ij},\Phi,\Pi_{\Phi},\Phi_{,i})$
and $T^{i}_{n}(h_{ij},\Phi,\Pi_{\Phi},\Phi_{,i})$, the general
Hamiltonian is derivable from the purely gravitational Hamiltonian
by means of
\begin{equation}
	\fbox{$
	\begin{array}{l}
		P^{A} \longrightarrow P^{A}+
		\sqrt{h}T^{i}_{n}y^{A}_{\,,i} \\
		\rho^{A}_{\,B} \longrightarrow \rho^{A}_{\,B}+
		2\sqrt{h}T_{nn}\delta^{A}_{\,B}
	\end{array}
	$}
\end{equation}
To be more specific, consider the case where $\Phi(x)$ stands for a
scalar field.
The corresponding energy/momentum projections are
\begin{equation}
	T_{nn} = \frac{1}{2}\left(\frac{1}{h}\Pi^{2}+
	h^{ij}\Phi_{,i}\Phi_{,j}\right) + V ~,~~
	T^{i}_{n} = \frac{1}{\sqrt{h}}\Pi h^{ij}\Phi_{,j} ~.
\end{equation}
In a more general case, e.g. for a gauge field $A_{\mu}$, the door
is open for non-gravitational constraints to enter the Hamiltonian.

\medskip
At the quantum level, we set ${\displaystyle P_{A}\equiv -i\frac
{\delta}{\delta y^{A}}}$.
Up to order ambiguities, the wave functional $\Psi$ of an empty 
brane-like Universe\cite{H,HH1,HH2} is subject to three Virasoro-type
constraints:
The momentum constraint equation
\begin{equation}
	y^{A}_{\,,i}\frac{\delta\Psi}{\delta y^{A}}=0 ~,
\end{equation}
is accompanied by the bifurcated Wheeler-Dewitt-like equation
\begin{equation}
	\fbox{$
	\begin{array}{l}
		{\displaystyle \frac{\delta}{\delta y^{A}}
	 	\left((\rho-\lambda I)^{-1}\right)^{AB}
	 	\frac{\delta}{\delta y^{B}}\Psi=\lambda\Psi}
		\vspace{4pt} \\
		{\displaystyle \frac{\delta}{\delta y^{A}}
	 	\left((\rho-\lambda I)^{-2}\right)^{AB}
	 	\frac{\delta}{\delta y^{B}}\Psi=\Psi}
	\end{array}
	$}
\end{equation}
Upon the inclusion of matter, the ordinary functional derivatives are 
replaced by covariant functional derivatives (and $\rho$ gets modified)
according to the above prescription.

\medskip
The Einstein limit of Regge-Teitelboim gravity has two faces:

\noindent $\bullet$ First, using the purely geometric relation
\begin{equation}
	2G_{nn}={\cal R}^{(3)}+
	\frac{1}{N^{2}}({\cal K}_{ij}{\cal K}^{ij}-{\cal K}^{2}) ~,
\end{equation}
we infer that
\begin{equation}
	\rho_{AB}-\lambda\eta_{AB} =
	2\sqrt{h}\left[
	(h^{ia}h^{jb}-h^{ij}h^{ab})y_{A\,|ab}y_{B\,|ij}+
	(G_{nn}-T_{nn})\eta_{AB}\right] ~.
\end{equation}
Appealing now to the embedding identity $\eta_{AB}y^{A}_{\,|ij}y^{B}
_{\,|k}=0$, one concludes that Einstein equation $G_{nn}=T_{nn}$
can be satisfied if and only if
\begin{equation}
	(\rho_{AB}-\lambda\eta_{AB})y^{B}_{\,|i}=0 ~.
\end{equation}
We have learned that the Einstein case is characterized by $\lambda$
being a $4$-fold degenerate eigenvalue of $\rho_{AB}$.
In turn, $(\rho-\lambda I)^{-1}$ does not make sense, and we face
the unpleasant consequence that not all components of $n^{A}$ are
expressible in terms of momenta.
This is, however, a curable situation.
The residual $n$'s are treated as non-dynamical variables, and the
troublesome $(\rho-\lambda I)^{-1}$ is replaced by some regular
$M^{-1}$, such that $ M^{-1}(\rho-\lambda I)$ defines the proper
projection operator. 

\noindent $\bullet$ Second, using the dynamical relation
\begin{equation}
	P_{A}=\sqrt{h}
	\left[(G_{nn}-T_{nn})n^{A}-(G_{ni}-T_{ni})h^{ij}y^{A}_{\,|j}+
	\left(y^{A}_{\,|j}n_{B}y^{B}_{\,|kl}(h^{ik}h^{jl}-h^{ij}h^{kl})
	\right)_{i}\right] ~,
\end{equation}
one observes that if Einstein equations $G_{ni}=T_{ni}$ and
$G_{nn}=T_{nn}$ are both satisfied, $P^{A}$ makes a total derivative.
On the other hand, reflecting the Poincare invariance of the embedding
spacetime, we know that the center-of-mass momentum $\mu^{A}\equiv\int
_{}^{}d^{3}xP^{A}$ is a Noether conserved vector.
And since the Arnowitt-Deser-Misner formalism exclusively involves
compact $3$-spaces, $\mu^{A}$ must vanish if Einstein equations are
to be respected.
Whereas a generic Regge-Teitelboim configuration exhibits a non-vanishing
Casimir $\mu^{2}=\eta_{AB}\mu^{A}\mu^{B}$, easily recognized as the
analogue of (mass)$^{2}$, Einstein configurations come with $\mu^{2}=0$.
In this language, Einstein gravity can be interpreted as the 'massless'
limit of Regge-Teitelboim gravity.

\end{document}